# Search-based Ordered Password Generation of Autoregressive Neural Networks

Min Jin, Junbin Ye, Rongxuan Shen and Huaxing Lu

*Abstract*—Passwords are the most widely used method of authentication and password guessing is the essential part of password cracking and password security research. The progress of deep learning technology provides a promising way to improve the efficiency of password guessing. However, current research on neural network password guessing methods mostly focuses on model structure and has overlooked the generation method. Due to the randomness of sampling, not only the generated passwords have a large number of duplicates, but also the order in which passwords generated is random, leading to inefficient password attacks.

In this paper, we propose SOPG, a search-based ordered password generation method, which enables the password guessing model based on autoregressive neural network to generate passwords in approximately descending order of probability. Experiment on comparison of SOPG and Random sampling shows passwords generated by SOPG do not repeat, and when they reach the same cover rate, SOPG requires fewer inferences and far fewer generated passwords than Random sampling, which brings great efficiency improvement to subsequent password attacks. We build SOPGesGPT, a password guessing model based on GPT, using SOPG to generate passwords. Compared with the most influential models OMEN, FLA, PassGAN, VAEPass and the latest model PassGPT in one-site test, experiments show that SOPGesGPT is far ahead in terms of both effective rate and cover rate. As to cover rate that everyone recognizes, SOPGesGPT reaches 35.06%, which is 254%, 298%, 421%, 380%, 81% higher than OMEN, FLA, PassGAN, VAEPass, and PassGPT respectively.

## I. INTRODUCTION

PASSWORDS, as a line of defense to ensure user information security, have the advantages of simplicity, flexibility, ease of deployment, and the ability to combine with other authentication technologies. They are currently the most widely used identity authentication method [1]. However, the use and storage of passwords are not always secure, and frequent password leakage incidents have increased people's attention and research on password security [2], [3]. The security of passwords is related to encryption algorithms, the computing power of attackers, and the password guessing methods used. As an indispensable and important component of password security research, password guessing is a key core technology in security attack and defense fields such as password strength assessment and password recovery [4], [5].

The essence of password guessing methods is to generate a set of passwords for dictionary attack. In the early stages of password generation, passwords were generated through dictionary and variant rule enumeration [6], [7]. This heuristic method requires manual formulation of transformation rules, lacks of theoretical basis, and the guessing effect depends on personal experience, often with significant randomness. Since 2009, a large number of well-known websites have experienced password leakage incidents [8], with a large number of passwords stored in plaintext, which not only sounded the alarm for password security but also provided real password samples for the research of password guessing methods. At this point, researchers proposed statistical password guessing methods represented by Markov models [9] and Probabilistic Context Free Grammar (PCFG) [10]. These two methods are theoretical and systematic, and the effect is better than generating passwords through variant rule enumeration. However, the characteristics of statistical models make it easy to overfit and fail to generate a large number of candidate passwords with good diversity.

The development of deep learning provides new approaches to improve password guessing technology. Theoretically, deep neural networks can spontaneously learn the structural features of password set and the interrelationship between characters, and billions of passwords brought by the frequent password leakage events help it more accurately depict the distribution of the real password set.

In 2016, Melicher et al. [11] first proposed a password guessing model based on Recurrent Neural Network (RNN) and achieved a higher cracking rate than PCFG and Markov

Min Jin, Rongxuan Shen are with Institute of Semiconductors, Chinese Academy of Sciences, Beijing 100083, China (e-mail: jinmin08@semi.ac.cn; rxshen@semi.ac.cn).
Junbin Ye was with Institute of Semiconductors, Chinese Academy of Sciences, Beijing 100083, China, and also with University of Chinese Academy of Sciences, Beijing 100089, China (e-mail: yejunbin@semi.ac.cn).
Huaxiang Lu is with Institute of Semiconductors, Chinese Academy of Sciences, Beijing 100083, China, with Materials and Optoelectronics Research Center, University of Chinese Academy of Sciences, Beijing 100049, China, and also with Beijing Key Lab of Semiconductor Neural Network Intelligent Perception and Computing Technology, Beijing 100083, China (e-mail: luhx@semi.ac.cn).

models. Since then, Password guessing models based on neural networks such as LSTM, GANs, and VAE have emerged like mushrooms after rain, demonstrating enormous technological potential. In 2017, Hitaj et al. [12] first proposed PassGAN, a password generation model based on GANs. In the experiment, the cracking rate of PassGAN increased by 18% to 24% compared to the automated password cracking tool HashCat. In 2020, Wang et al. [13] first proposed the VAE based password generation model PGVAE, which achieved better cracking rates than PassGAN in both same-site and cross-site attacks. Later, Yang et al. [14] proposed VAEPass, a lightweight VAE password guessing model with better effect. In 2022, Ye et al. [15] first proposed a password guessing model based on TCN, and the test results were superior to OMEN, PassGAN, and VAEPass. Transformer (or GPT, a transformer-based language model) is also used in password guessing [16], [17], [18]. The self attention mechanism in Transformer allows it to better focus on the internal connections between passwords, and gain better effect than the GAN model and VAE model in password guessing. Scholars also combine neural network models with traditional models to further improve the efficiency of password guessing. The PL model proposed by Liu et al. [19] combines PCFG and LSTM, takes the basic structure of the password as the network input, and uses LSTM to predict the following basic structure. This model improves the password training from character-based to word-based, thereby significantly improving the hit rate of a single dataset. Wang Ding et al. [20] combined RNN with PCFG and proposed models such as PR and PR+, whose cracking rate was significantly higher than that of a single model.

Overall, current research on neural network password guessing methods mostly focuses on model structure. Through the design of the model structure, the modeling ability of the password dataset is improved, thereby improving the success rate of password guessing. But there is a problem that has been overlooked, that is, the generation method of neural network password guessing models. Currently, there are three main types of neural networks used for password guessing models: autoregressive networks (such as Transformer, LSTM, TCN), generative adversarial networks (GAN), and variational autoencoders (VAE). These models generate candidate passwords based on the learned distribution features during the generation stage. Due to the randomness of sampling, not only there are a large number of repeated passwords, resulting in a waste of computing resources, but also the order in which passwords are generated is random, and there is no guarantee that passwords with higher probability will be generated first. As we know, password guessing is for password attacks. The earlier a password with high probability is tried, the sooner it can be successfully cracked and the calculation efficiency will be improved. The reason why the Ordered Markov Enumerator (OMEN) [21] is widely used also is that it can generate candidate passwords in an approximate probability descending order, greatly improving the efficiency of subsequent dictionary attacks.

The GAN based password guessing model generates a candidate password from a random noise, while the VAE based model samples from a Gaussian distribution, making it difficult for them to generate passwords in probability descending order. The model based on Transformer and LSTM adopts an autoregressive form, and the generation of the current character depends on the previous characters, using Monte Carlo random sampling. OMEN also uses autoregressive form to generate passwords. Based on the n-grams Markov model, the probability of the next character appearing depends on the first n characters. OMEN can calculate the probabilities of all n-grams structures in advance, discretize all probabilities into many integer intervals, and assume that the password length is controllable, enumerate to generate passwords on all these intervals in descending order of probability. The length of the password generated by an autoregressive neural network is uncontrollable, and the probability of the next character depends on the previous character, making it impossible to calculate the probability in advance. Therefore, the idea of OMEN cannot be generalized and applied to the autoregressive neural network password guessing models.

Unlike natural language processing tasks that only require a single high-quality solution, password guessing necessitates the generation of extremely large candidate passwords. Compared to Monte Carlo random sampling, search algorithms can avoid repeated password inference and are more suitable for generating passwords. Wang Ding et al. [20] used breadth first search with minimum boundary constraints when generating passwords, which can generate passwords greater than a certain probability threshold. However, this article mainly focuses on the model structure and does not discuss the generation methods too much.

Theoretically, using priority queue-based breadth-first search can achieve descending order in probabilities. However, the character space of a password contains a total of 95 letters, numbers, and special characters; and the length of passwords is usually between 6 and 16. This means if the password space is considered as a probability tree, the node size of the tree reaches $95^{16}$ (approximately $4.4 \times 10^{31}$), which cannot be supported by the memory. Therefore, the probability descending generation of the autoregressive neural network password models remains a challenging problem that hasn't been addressed in existing work.

In this article, we propose SOPG, a search-based ordered password generation method, which enable the password guessing model based on autoregressive neural network to generate passwords in approximately descending order of probability. A password guessing model based on GPT was built to support SOPG. We choose GPT as a representative of autoregressive neural network because its performance in password guessing is superior to other autoregressive neural network models such as LSTM.

The core ideas of SOPG includes three points: Firstly, a method of packing low probability nodes is proposed to optimize search nodes and solve the storage problem caused by node exponential explosion; Secondly, a frontier, "PGQueue", is designed to ensure that the max number of nodes is limited to around *N* during the entire search process, and expand the node with the highest probability as much as possible each time, which make passwords with higher probability be generated first. Thirdly, Concurrent multiple sub-searches to relieve search trapped in local optima. SOPG enables the current password guessing model based on autoregressive neural network to generate passwords in approximately descending order of probability, greatly improving the success rate and computational efficiency of subsequent password attacks, making neural network password guessing a big step towards practical applications.

Our main contributions are summarized as follows:
- **The first ordered password generation method of neural network.** We proposed SOPG to generate passwords in approximately descending order of probability for autoregressive neural networks. Compared to the currently used random sampling, passwords generated by SOPG do not repeat, and when they reach the same cover rate, SOPG requires fewer inferences, which means SOPG generates passwords with higher efficiency. Moreover, when reaching the same cover rate, random sampling needs to generate far more passwords than SOPG, which means SOPG brings great efficiency improvement to the subsequent password attacks.
- **A state-of-art password guessing model in one-site tests.** We build SOPGesGPT, a password guessing model based on GPT, using SOPG to generate passwords. Compared with the most influential models OMEN, FLA, PassGAN, VAEPass and the latest model PassGPT in one-site test, our model has the state-of-art performance and improves the cover rate to 35.06% in the test, which is 254%, 298%, 421%, 380%, 81% higher than OMEN, FLA, PassGAN, VAEPass, and PassGPT respectively.

The rest of this article is organized as follows. Section II briefly introduces the related work. Section III presents the principle and design of SOPG. Section IV evaluates the effectiveness and efficiency of SOPG. We conclude the article and discuss further work in Section V.

## II. RELATED WORK

### A. Neural network generative models

Neural network generative models generally use maximum likelihood. Some generative model not using maximum likelihood, but can also be expressed in the form of using maximum likelihood (such as GAN) [22]. For a dataset containing m training sample $x^{(i)}$, define a model with a likelihood of $\prod_{i=1}^{m} p_{model}(x^{(i)}; \theta)$, $\theta$ is the parameter of the model. Maximum likelihood is to select parameter $\theta$ that maximize the likelihood of training data for the model. The neural network generative model can be divided into several types as shown in Fig.1 through the maximum likelihood principle or approximation method.

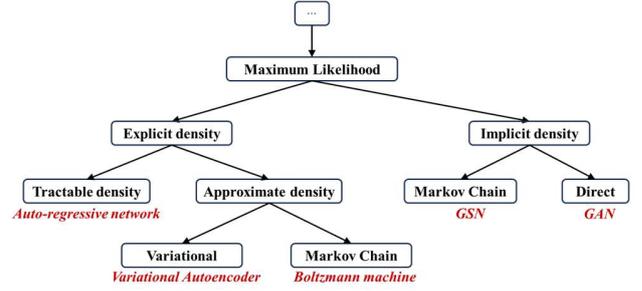

**Fig. 1.** This is the category of neural network generative models.

The left side of Fig.1 called Explicit density model defines the density distribution of the model $p_{model}(x^{(i)}; \theta)$ explicitly. For such models, we bring the density distribution into the likelihood expression, and then update the model along the ascending direction of the gradient to calculate the maximum likelihood. There are roughly two ways:
- Tractable explicit models: These models are called Auto-regressive networks. They define a density distribution easy to calculate and transform the likelihood into the joint product form of conditional probability through the chain rule of probability [23].
- Approximate explicit models: They can be divided into Variational Autoencoder (VAE) using variational method and Stochastic Approximation using Markov chain Monte Carlo method, such as Boltzmann machine.

The right side of figure 1 called Implicit density models can be trained without explicitly defining density functions. Simultaneously, they provide a method for training models only indirectly interacting with $p_{model}$, usually by sampling from $p_{model}$. They can also be divided into two categories:
- Generative stochastic network (GSN) [24]: It takes samples from the Markov chain after reaching a stationary distribution.
- Generative adversarial network (GAN): It is a generative model training through mutual game between two neural networks. After the training is completed, it generates a sample in one step using random noise as input.

### B. Password generative model of Auto-regressive neural network

Password guessing models based on different neural networks have different ways of generating passwords. As described in II.A, the password generative model based on GAN is sampled from random noise; the password generative model based on VAE is sampled from Gaussian distribution. The password generative model based on RNN, LSTM, or Transformer adopts the form of autoregression; the current character generation relies on the previous character and generates the next character through random sampling using the Monte Carlo method. Fig.2 shows an example of RNN generating password. All these password generative models cannot guarantee the priority generation of passwords of high probability due to the randomness of sampling, which seriously affects the computational efficiency of subsequent password attacks.

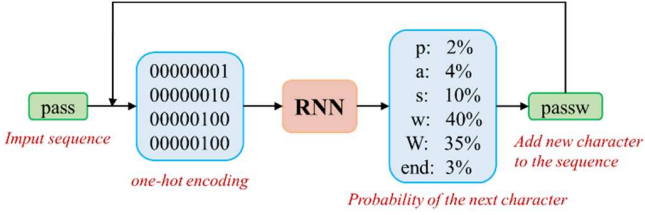

**Fig. 2.** This is an example of Autoregressive neural network generating password.

*C. OMEN*

Ordered Markov Enumerator (OMEN) based on the Markov n-grams model can generate password candidates according to their occurrence probabilities, i.e., it outputs most likely passwords first. On a high level, it discretizes all probabilities into a number of bins, and iterates over all those bins in order of decreasing likelihood. For each bin, it finds all passwords that match the probability associated with this bin and outputs them. More precisely, it first takes the logarithm of all n-gram probabilities, and discretizes them into levels, according to the formula $lvl_i = round(\log(c_1 \cdot prob_i) + c_2))$, where $c_1$ and $c_2$ are chosen such that the most frequent n-grams get a level of 0 and that n-grams that did not appear in the training are still assigned a small probability. It adjusts the parameter $nbLevel$ to get the desired number of levels ($nbLevel$), i.e., the levels can take values 0, -1, …, $-(nbLevel - 1)$, where $nbLevel$ is a parameter. The number of levels influences both the accuracy of the algorithm as well as the running time: more levels mean better accuracy, but also a longer running time.

Although OMEN uses autoregressive form to generate passwords the same as the autoregressive neural networks, its approach cannot be generalized. Because, OMEN needs to calculate the probabilities of all n-grams structures in advance and assume that the password length is controllable, while the length of the password generated by an autoregressive neural network is not controllable, and the probability of the next character depends on the previous characters making it impossible to calculate the probability in advance.

*D. Search Algorithm*

Search refers to the process of finding a sequence of actions from one state to another in order to achieve the goal of solving a problem. The state space of a problem is a set of all states that can be reached from the initial state of the problem, and the states in the state space are called nodes. If a new node can be expanded from the current node, the current node is called the parent node, and the expanded new node is the child node of the current node, and nodes without children are called leaf nodes. At any given moment, the set of all leaf nodes to be extended is called frontier. The search algorithm is the process of selecting nodes in the frontier and expanding new nodes, repeating this process until the solution is found or there are no nodes in the frontier that can be extended. The basic structure of different search algorithms is roughly the same, with the main difference being how to select the nodes to be extended.

Tree search ensure that the state does not repeat during the search process. The length of passwords in the autoregressive generation mode is getting longer and longer, so the search path will not repeat. Therefore, Tree search can be used for search-based password generation of the autoregressive neural networks.

The search strategy of Breadth-First Search (BFS) first expands the root node, then expand all the children of the root node, and then expand the children of all the children of the root node, and so on. If the path cost is a non-decreasing function based on node depth, BFS is optimal. But BFS consumes a huge amount of time and space. Assuming each node has b child nodes and the depth of the solution is d, the time complexity is $O(b^d)$ and the space complexity is $O(b^{d-1})$, which both have exponential complexity. If the maximum length of the password is 16 and the allowed characters of the password are 95 printable ASCII characters, the space complexity of BFS for search-based password generation is unbearable $O(95^{15})$, and cannot generate passwords in descending order of probability.

For arbitrary path cost, the first expanded node is not necessarily optimal, and Breadth-first search is no longer the optimal solution. Uniform-Cost Search (UCS) sorts the nodes in the frontier according to path cost, and expands the node with the lowest cost each time. Using $C^*$ to represent the cost of the optimal solution, and assuming the cost of each action is at least $\varepsilon$, the time and space complexity of uniform cost search is $O\left(b^{1+\lceil\frac{C^*}{\varepsilon}\rceil}\right)$ in the worst case. Take the probability of generating a password as the path cost, UCS can be used for search-based password generation. Assuming the maximum length of the password is 16, the time and space complexity of password generation by UCS are both $O(95^{15})$. The generated passwords by UCS are arranged in descending order, but the huge space cost is still unacceptable.

Depth-First Search (DFS) always extract the deepest node from the frontier for expansion. Assume that the maximum depth of any node is $m$, in the worst case, the time complexity of DFS is $O(b^m)$. DFS only needs to store the path from the root node to the current node, that is, the space complexity is $O(bm)$. Assuming the maximum length of the password is 16, the space complexity of password generation by DFS is $O(95 * 12)$, which is greatly reduced compared to BFS and UCS. However, DFS cannot generate passwords in descending order of probability.

III. SEARCH BASED ORDERED PASSWORD GENERATION OF AUTOREGRESSIVE NEURAL NETWORKS

Motivated by OMEN and Search algorithm, we present a search-based ordered password generation method for autoregressive neural networks named SOPG, which enable the password guessing model based on autoregressive neural networks to generate passwords in approximately descending order of probability. Meanwhile, we build SOPGesGPT, a password guessing model based on GPT, to support SOPG.

*A. Overall Architecture*

Fig.3 shows the framework of SOPG. Firstly, we preprocess the password dataset and split it into training set and testing set. Next, encode the password and train a password guessing model based on autoregressive neural network. Then, generate password candidates using search-based ordered password generation method which is the main contribution of this article. During the experimental phase, the new password set can be tested by the testing set.

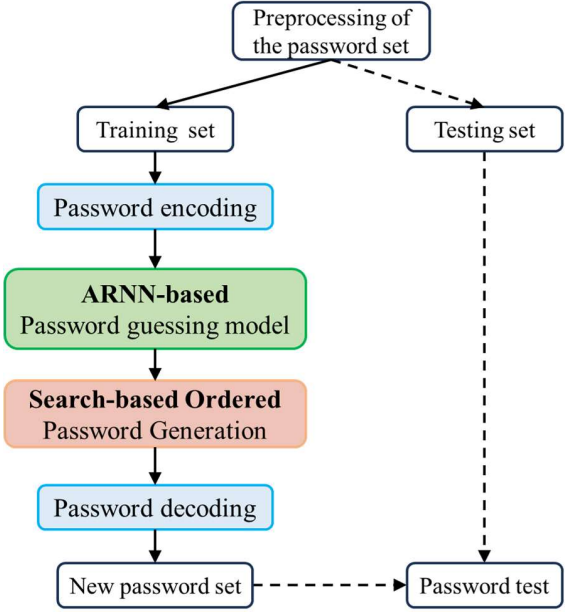

Fig.3. The framework of SOPG.

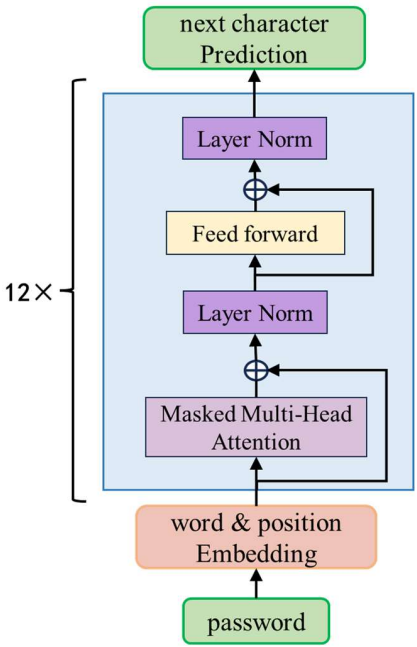

Fig.4. The structure of SOPGesGPT.

*B. Password Guessing Model based on GPT*

Due to the excellent performance of transformer, we choose GPT, a transformer-based language model, as an example of autoregressive neural networks to build a password guessing model, which is called SOPGesGPT. The structure of SOPGesGPT is shown in Fig.4 and the parameter settings are as follows:
- Set the number of self-attention layers, i.e., the number of model layers, to 12 layers.
- The feature dimension of the word vector is set to 512.
- Set the number of heads for mask multiple head attention to 8.
- The dimension of the hidden layer of feedforward neural network is set to 1024.
- Use learnable position encoding.

*C. Password Encoding*

In order for computers to recognize different characters, it is necessary to encode the original password before sending it to the model. Except for all 95 printable characters in the ASCII set, we additionally use 4 special identifiers: <S>, </S>, <unk> and <blank>. <S> represents the beginning of a password. </S> represents the end of a password. <unk> is used to represent unknown characters that are not within the encoding range. <blank>, the blank placeholder, is used to fill in the passwords, due to the length of each password will increase to the length of the longest password in the current batch. Specific recognizable characters of SOPGesGPT are shown in Table 1.

TABLE I
CHARACTER CLASSIFICATION FOR SOPGESGPT

| character classification | Specific characters included |
| --- | --- |
| 52 English characters | ABCDEFGHIJKLMNOPQRSTUVWX YZabcdefghijklmnopqrstuvwxyz |
| 33 special characters | `~!@#$%^&*()_-+=[\]{|};',./:"<>? |
| 10 numeric characters | 0123456789 |
| 4 special identifiers | <S>, </S>, <unk>, <blank> |

Assuming the maximum input length allowed by the model is $n$, and the encoding function for passwords is $encode()$, which maps characters one by one to integers. For a password $code = C_1C_2\cdots C_m$ with a length of $m(m < n)$, the corresponding input sequence of the GPT password guessing model is: $X = \{x_1, x_2, \cdots, x_n\}$, where $x_i$ is got through Eq.1.

$$x_i = \begin{cases} encode(<S>) & if\ i = 1 \\ encode(C_i) & if\ 1 < i \le m+1 \\ encode(<blank>) & if\ m+1 < i \le n \end{cases} \quad (1)$$

The corresponding output sequence of the GPT password guessing model to $code = C_1C_2\cdots C_m$ is: $Y = \{y_1, y_2, \cdots, y_n\}$, where $y_i$ is got through Eq.2.

$$y_i = \begin{cases} encode(C_i) & if\ 1 \le i \le m \\ encode(</S>) & if\ i = m+1 \\ encode(<blank>) & if\ m+1 < i \le n \end{cases} \quad (2)$$

Note that there is a part of the input sequence $X$ which is the same as the output label $Y$, with only one bit staggered. It is through this dislocation that the generation probability of the next character is predicted.

*D. Search-based Ordered Password Generation*

A large number of passwords must be generated through a trained password guessing model. There are two disadvantages of using Monte Carlo random sampling: First is that a large number of repeated passwords will be generated; the second is that it cannot generate passwords in descending order of probability, which is beneficial to the subsequent password attacks. Therefore, we proposed the Search-based Ordered Password Generation (SOPG). In this section, we will explain the core idea of SOPG.

As we mentioned in III.C, the input of the GPT password guessing model can be expressed as $X = \{x_0, x_2, \cdots, x_n\}$, where $n$ is the maximum length of password we set. Through this encoding, each input state of the model can be characterized making it possible to search, and each input state of the model can be represented in Eq.3, in which, $X$ is the input of the model, $deep$ represents the character number at which the model is referring, and after it, there are all blank placeholders. $log\_prob$ performs logarithmic operation on the probability of input string, due to the logarithmic probability ensures that probability multiplication does not lose accuracy.

$$State := (X, deep, log\_prob) \quad (3)$$

After defining the input state of the model, we can search to generate passwords. As we mentioned in II.D, the existing search strategies, such as BFS, UCS and DFS, are not applicable for password generation, due to the unbearable space complexity or uncontrolled order of probability. We proposed a new search strategy for password generation call SOPG.

SOPG has optimized the search nodes. The allowed characters of the password are 95 printable ASCII characters, and the score distribution of these characters is uneven when the scoring function be calculated at each step of the model. With a specific character prefix, only a small number of characters have high scores, while the scores of others are very low. This will result in adding too many low probability nodes to the frontier, wasting a lot of space. In SOPG, we proposed a method of packing low probability nodes to optimize the space. First of all, set a packing probability $P_0$. When searching a node, expand normally for all nodes with a probability value bigger than $P_0$, and pack all nodes with a probability value smaller than $P_0$ into a new node, as shown in Fig.5. When a packed node is searched again, expand it to nodes with probability value smaller than $P_0$ and bigger than $P_1$, meanwhile, pack all nodes with a probability value smaller than $P_1$ into a new node (where $P_1 < P_0$), as shown in Fig.6. We set a series of descending parameters $P_0, P_1, P_2, \cdots, P_s$ for stepwise expanding the packed nodes. To save space, the packed node only stores the state of the parent node, and do not store the state of the expanded child nodes. When expanding a packed node, it needs to re-infer the stored parent node and return nodes less than the packing probability.

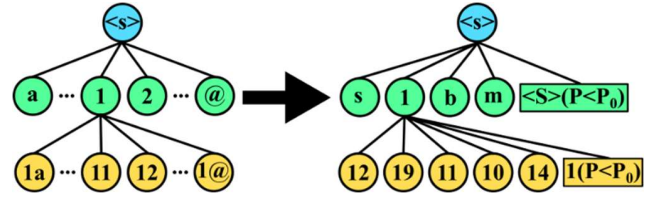

Fig.5. Diagram of packing search nodes. At this point, the node being searched is <S>1. Our method packs all child nodes of <S>1 with a probability value smaller than $P_0$ into a new node <S>1$(P < P_0)$.

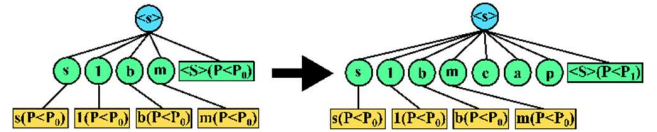

Fig.6. Diagram of expanding a packed search node. At this point, the node be expanded is <S>$(P < P_0)$. Our method needs to re-infer the node <S> and return child nodes with a probability value smaller than $P_0$ and bigger than $P_1$, meanwhile, packs all child nodes with a probability value smaller than $P_1$ into a new node <S>$(P < P_1)$, where $P_1 < P_0$.

In SOPG, we designed a frontier "PGQueue" that meets the following requirements:

1): The number of nodes in the frontier is limited to $N$.

2): The state of a search node is expressed as $(input, deep, prob, threshold)$, where $input$ represents the current password; $deep$ represents the current search position, i.e., the depth of the search tree; $prob$ represents the logarithmic probability of the current password, that is, the search path cost; $threshold$ is the packing probability of search node optimization, if it is non-zero, it means that the node is a packing node, and it needs to re-infer the stored parent node and return nodes with probability less than $threshold$ when expanding it.

3): The starting node of the frontier is $(input_0, 0, 0, 0)$, the first of $input_0$ is the encoding of the beginning character <S>, the rest are the encoding of the blank placeholder <blank>.

4): The state comparison of the search node: first compare the value of $prob$ of the two states, and then compare the value of $threshold$.

5): According to different values of $deep$, maintain several independent maximum heaps. The search nodes with the same value of $deep$ are in the same maximum heap, and the number of the maximum heaps depends on the maximum length allowed by the password. The purpose of using the maximum heap is to ensure that new nodes can be added or the largest node at this depth can be taken out with the complexity of $\log m$, $m$ is the number of all nodes with the same search depth.

6): Add new nodes (push): Put it into the corresponding maximum heap according to the value of $deep$.

7): Fetch nodes (pop): When the number of nodes in the frontier is less than $N$, take out the node at the top of the maximum heap whose top node has the largest $prob$. When the number of nodes in the frontier is more than or equal to $N$, take out the node at the top of the maximum heap with the largest value of $deep$.

Our designed frontier, "PGQueue", can ensure that the max number of nodes is limited to around $N$ during the entire search process, and expand the node with the highest probability as much as possible each time. We use frontier "PGQueue" for search, when the number of nodes in the frontier is less than $N$, it is similar to Uniform-Cost Search; when the number of nodes in the frontier is greater than or equal to $N$, it is equivalent to perform Depth-First Search on the nodes with the longest password length. This is a balance we designed due to the space constraints, which makes the generated passwords not strictly in descending order of probability, but in approximate descending order of probability. SOPG avoids local optima by concurrent multiple sub-searches, and this strategy can also improve hardware utilization by combining nodes of different sub-searches into a batch and sent to SOPGesGPT together. Algorithm 1 shows the details of SOPG.

## IV. EXPERIMENT AND RESULTS

### A. Dataset and evaluation metric

The dataset used in this paper is the password dataset leaked by RockYou, a famous social networking website, in 2009. There are 32603388 plaintext passwords in this password dataset. Considering that the characters used in passwords are generally printable ASCII characters, we cleaned up 18038 unqualified passwords. Also, the password length requirement of the websites is probably between [6,32], so we clean 1407508 other passwords beyond the range. Finally, the dataset resulted in a total 31177842 passwords, and the number of unique passwords is 14048038. Hence, the password repetition rate of the RockYou dataset is about 54.9%.

We use cover rate and effective rate as the evaluation metric for the quality of the generated passwords, the cover rate and effective rate are respectively defined in (4) and (5):

$$Cover\ Rate = \frac{cnt(New_u \in (Test - Train))}{cnt(Test - Train)} \times 100\% \quad (4)$$

$$Effect\ Rate = \frac{cnt(New_u \in (Test - Train))}{cnt(New_u)} \times 100\% \quad (5)$$

Where $New_u$ represents the unique generated passwords, $(Test - Train)$ means the unique passwords in the testing set and not in the training set. $cnt()$ counts the number of elements in the set. In addition, we will also discuss the space consumption and the number of inference times of SOPG when generating passwords.

We randomly divide the RockYou dataset into a training set and a testing set according to the ratio of 8:2. We train the model with the training set and compare the result with the testing set to calculate the generated passwords' cover rate. When the SOPGesGPT is trained, the "dropout" is set to 0.1, the largest length of passwords is set to 32, and batch size is 128.

---

**Algorithm 1** Search-based Ordered Password Generation

**Input:** SOPGesGPT, a trained password guessing model based on GPT. Search threshold $P_{min}$, the lower bound of the generated passwords' probabilities. $P_0, P_1, P_2, \cdots, P_s$, a series of descending parameters for packing search nodes. The total frontier Q and frontier $Q_i$ of each sub-search, adopting our designed frontier "PGQueue" and the number of nodes is limited to $N$.

**Output:** Ordered passwords with probabilities bigger than $P_{min}$.

1: **while** true **do**
2:    node ← return node and remove it from Q
3:    feed node into SOPGesGPT for inference
4:    **for all** child nodes expanded based on the inference results of SOPGesGPT **do**
5:      **if** the node's probability < $P_{min}$ **then**
6:        delete the node
7:      **else if** the node end with </s> **then**
8:        output the node as a password
9:      **else** push the node into Q    **end if**
10:   **end for**
11:   **if** the number of nodes in Q ⩾ $N$ **then**
12:      end the while loop
13:   **end if**
14: **end while**
15: Create m sub-search processes, the frontier $Q_i$ of each sub-search process is independent and initialized to empty.
16: **for all** sub-search processes **do**
17:   **while** true **do**
18:     **if** $Q_i$ is empty **then**
19:       **if** Q is empty **then**
20:         end the while loop of the i-th sub-search
21:       **end if**
22:       fetch k nodes from Q and add to $Q_i$
23:     **end if**
24:     fetch a node from $Q_i$ for inference
25:     **for all** child nodes expanded based on the inference results of SOPGesGPT **do**
26:       **if** the node's probability < $P_{min}$ **then**
27:         delete the node
28:       **else if** the node end with </s> **then**
29:         output the node as a password
30:       **else** push the node into $Q_i$    **end if**
31:     **end for**
32:   **end while**
33: **end for**

## B. Search threshold $P_{min}$

Search threshold $P_{min}$ determines the number of generated passwords, in this section, we will explore the influence of $P_{min}$ through experiments. Specifically, we use the same trained password guessing model based on GPT(SOPGesGPT) and set different search thresholds, generate passwords using SOPG and calculate cover rate and effective rate according to Eq.4 and Eq.5 under different search thresholds. The experimental data is shown in Table I, the Match Number denotes the number of unique new passwords in the testing set, and the Hit Number denotes the number of unique new passwords in the testing set and not in the training set.

TABLE II
COVER RATE AND EFFECTIVE RATE UNDER DIFFERENT SEARCH THRESHOLDS $P_{min}$

| Search Threshold | $New_u$ | Match Number | Hit Number | Cover Rate | Effect Rate |
|---|---|---|---|---|---|
| $1.0 \times 10^{-7}$ | 589,808 | 409,360 | 9,492 | 0.40% | 1.61% |
| $1.0 \times 10^{-8}$ | 4,914,678 | 974,112 | 219,924 | 9.19% | 4.47% |
| $5.0 \times 10^{-9}$ | 9,219,721 | 1,150,849 | 335,558 | 14.02% | 3.64% |
| $1.0 \times 10^{-9}$ | 44,614,795 | 1,586,321 | 672,352 | 28.10% | 1.51% |
| $5.0 \times 10^{-10}$ | 78,300,418 | 1,740,255 | 798,866 | 33.39% | 1.02% |
| $4.0 \times 10^{-10}$ | 94,740,119 | 1,788,172 | 838,929 | 35.06% | 0.89% |
| $3.0 \times 10^{-10}$ | 122,846,669 | 1,851,724 | 892,550 | 37.30% | 0.73% |

As we noticed from Table II, the Hit Number is smaller than the Match Number in each row, especially when search threshold $P_{min} = 1.0 \times 10^{-7}$, the Hit Number is far smaller. That is because many of the generated passwords have appeared in the training set, which is an inherent problem in password guessing using deep learning technology. Given that our SOPG can generate passwords in approximately descending order, this proportion is higher in the early stage of password generation, but as the number of generated passwords increases, the proportion will decrease. This is also reflected in Table II.

From Table II, as the search threshold $P_{min}$ decreases, more passwords are generated, the cover rate also increases accordingly, meanwhile, the effective rate of the generated passwords is decreasing. The first row of data in the table deviates slightly from the pattern and the reason is as mentioned in the previous paragraph. By presenting the data in graphics, as shown in Fig.7 and Fig.8, we can see it more intuitively. This means in the roaming attack, to achieve a high cracking rate (i.e., cover rate here), the computational efficiency of password attacks in the later stages will be very low. Of course, this is also a problem with other password guessing algorithms, and in subsequent comparisons with other algorithms, it can be found that our effective rate is still significantly higher than other algorithms.

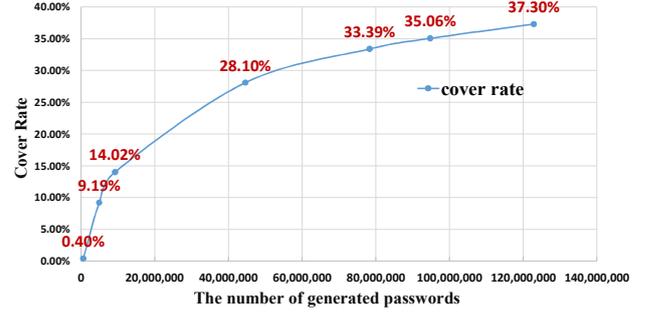

Fig.7. Relationship between generated password number and cover rate.

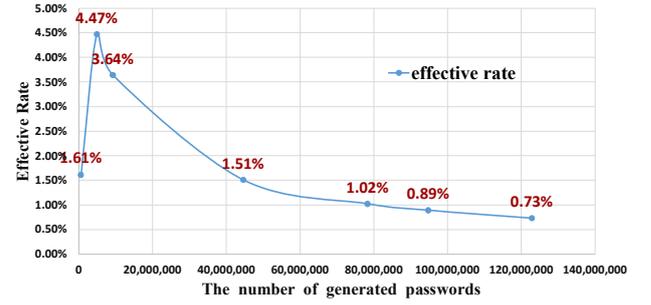

Fig.8. Relationship between generated password number and effective rate.

## C. Space Saving and the Setting of PGQueue' N

In order to analyze the advantages brought by SOPG on space and the setting of PGQueue' $N$, this section sets up the following experiments: based on the same trained password guessing model (SOPGesGPT), generate passwords using search methods and stop searching until the number of nodes in the frontier reach to $N$; compare the number of passwords that different search methods have generated. The search methods for comparison include:

- **Uniform-Cost Search (UCS):** take the probability of generating a password as the path cost and the frontier adopts a priority queue.
- **Uniform-Cost Search + pruning:** add pruning to UCS, pruning threshold is set to $10^{-9}$, that is, nodes with a probability less than $10^{-9}$ are not added to the frontier, thereby reducing the size of the search space.
- **SOPG (without sub-searches process):** pruning threshold is also set to $10^{-9}$.

Use the above three algorithms to perform password generation on the same trained SOPGesGPT, compare the number of passwords that can be found at different $N$. The experimental data is shown in Table III.

From Table III we can notice that, regardless of the capacity of the frontier ($N$), SOPG can search for far more passwords than UCS+Pruning, which means that for the same number of passwords generated, SOPG can save a lot of space. It is worth mentioning that we should not set too big $N$, because as the number of nodes in the fronter increases, the calculations required to extract the node from frontier also increase, and in turn reduce the inference times per second and affect the speed of password generation.

#### TABLE III
THE NUMBER OF PASSWORDS CAN BE SEARCHED BY DIFFERENT SEARCH ALGORITHMS UNDER THE LIMITED FRONTIER LENGTH

| $N$ | UCS | UCS+Pruning | SOPG |
| --- | --- | --- | --- |
| 10,000 | 1 | 1 | 523 |
| 100,000 | 14 | 23 | 9,196 |
| 1,000,000 | 734 | 1,701 | 133,188 |
| 10,000,000 | 13,426 | 69,474 | 2,068,810 |

### D. Comparison of SOPG and Random sampling

SOPG performs packing operations on low probability nodes to save space. However, when expanding a packed node, it needs to re-infer the stored parent node which increases the calculation time and the number of model inferences. In this section, we will compare SOPG with random sampling, the mainstream of neural network password generation. Use random sampling and SOPG to perform password generation on the same trained SOPGesGPT. Use (4) to calculate the cover rate, when the cover rate reaches 0.4%, 9.19%, 14.02%, compare the number of passwords that need to be generated, the number of non-repeating passwords, and the number of inferences between random sampling and SOPG. The number of model inferences are calculated as follows:

- **Random sampling:** the number of inferences required to generate a password is equal to the length of the password plus one.
- **SOPG:** record the number of inferences of the model during the process of searching for passwords as the actual number of inferences.

The experimental data is shown in Table IV, RS is used to refer to random sampling.

#### TABLE IV
COMPARISON OF MODEL INFERENCE TIMES OF RANDOM SAMPLING AND SOPG

| Cover Rate | Methods | Generate Password | Unique Password | Inference Times |
| --- | --- | --- | --- | --- |
| 0.40% | RS | 1,076,923 | 704,391 | 9,669,394 |
|  | SOPG | 589,808 | 589,808 | 5,382,789 |
| 9.19% | RS | 45,141,507 | 19,673,319 | 405,257,615 |
|  | SOPG | 4,914,678 | 4,914,678 | 39,826,710 |
| 14.02% | RS | 96,709,827 | 38,525,882 | 771,491,765 |
|  | SOPG | 9,219,721 | 9,219,721 | 73,266,252 |

From Table IV we can notice that, random sampling results in a large number of repeated passwords, while passwords generated by SOPG are not repeated; and when they reach the same cover rate, SOPG requires fewer inferences. In the early stage of password generation (the cover rate reaching 0.40%), the number of inference times required for SOPG is 55.7% of random sampling. As the cover rate increases to 14.02%, the number of inference times required for SOPG is only 9.5% of random sampling. In order to see it more intuitively, we present the data in graphics, as shown in Fig.9 and Fig.10. From Fig.10, we can see that when reaching the same cover rate, the passwords RS needs to generate are much more than SOPG, which means SOPG will bring great efficiency improvement to subsequent password attacks.

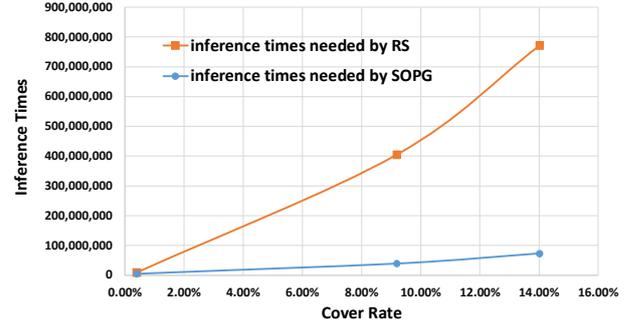

Fig.9. Inference times needed by Random sampling and SOPG under different cover rate.

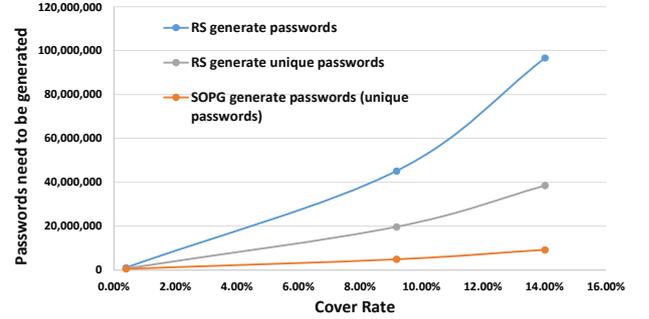

Fig.10. Passwords need to be generated by Random sampling and SOPG under different cover rate.

### E. Quality of generated passwords of different models

To further validate the performance of SOPG, the comparison with other password guessing models is conducted in this section. We select five of the most influential models for comparison, and the details of each model are as follows:

- **SOPGesGPT:** our proposed model, using SOPG to generate passwords, whose parameters is described in III.B.
- **OMEN:** use the code in [26] to run a 4-ngram OMEN with the training set we used.
- **FLA:** Since the dataset used in [11] is a relatively small specific testing dataset, we train a 3-layer FLA under the dataset we used to compare with other models better. The hidden size of LSTM is 256, and we use the dropout layer between each LSTM layer. We train the FLA for 20 epochs at batch size 512.

- **PassGAN:** The dataset segmentation ratio of RockYou used by PassGAN [12] is the same as 8:2 in this paper, so we use the result in [12].
- **VAEPass:** We instantiate the model in [14], where the core module for constructing the VAE encoder and decoder is the GCNN (Gated Convolutional Networks [25]). The number of encoders and decoders is 2 and 1, respectively. Other parameters of the model are input sequence length of 12, hidden layer of 256, latent space of 64, and convolution kernels of 5. We train VAEPass for 20 epochs with a batch size of 128.
- **PassGPT:** The dataset segmentation ratio of RockYou used by PassGPT [27] is also 8:2 as in this paper, so we use the result in [27]. In [27], only the Cover Rate is provided directly. It also provided the percentage of unique passwords under different number of guesses, and we can calculate that the number of unique passwords is 80,000,000 when the number of guesses is $10^8$. There is no information in [27] indicating the Hit Number and Effect Rate, so these two grids in Table V are empty.

Use the above five models to generate passwords separately, and compare cover rate and effective rate. For OMEN, FLA, PassGAN, VAEPass and PassGPT, the number of generated passwords is set to $10^8$, and number of unique passwords is recorded. As to SOPG, set search threshold $P_{min}$ to $4.0 \times 10^{-1}$ and the number of passwords generated is 94740119, which is less than $10^8$, but on the same scale. The setting of the search threshold can be determined by generating a larger number of passwords, and then make the number of generated passwords accurately equal to $10^8$, but it is not necessary. In practical applications, if the number of generated passwords is not enough, the search threshold can be appropriately reduced, referring to IV.B, and set the search upper bound to avoid repeat search.

TABLE V
THE PERFORMANCE OF SOPGESGPT COMPARED WITH OTHER MODELS WHEN $10^8$ PASSWORDS ARE SAMPLED

| Models | Unique Passwords | Hit Number | Effect Rate | Cover Rate |
|---|---|---|---|---|
| SOPGesGPT | 94,740,119 | 838,929 | 8.86‰ | 35.06% |
| OMEN | 100,000,000 | 234,594 | 2.35‰ | 9.89% |
| FLA | 68,655,732 | 211,077 | 3.07‰ | 8.82% |
| PassGAN | 52,815,412 | 133,061 | 2.52‰ | 6.73% |
| VAEPass | 62,064,655 | 160,589 | 2.59‰ | 7.30% |
| PassGPT | 80,000,000 | / | / | 19.37% |

Table V shows the comparison results of the six models. SOPGesGPT and OMEN do not generate duplicate passwords, so the number of unique passwords equals the number of passwords generated. We can notice that the password generated by FLA, PassGAN and VAEPass have a large number of duplicates, while PassGPT is slightly better. As to FLA, PassGAN, VAEPass and PassGPT, we deduplicate the passwords generated and record the number of unique passwords. We use the number of unique passwords as denominator to calculate effect rate, which is beneficial for FLA, PassGAN, VAEPass and PassGPT. From Table V, we can see that our model SOPGesGPT is far ahead in terms of both effective rate and cover rate. As to cover rate that everyone recognizes, SOPGesGPT reaches 35.06%, which is 254%, 298%, 421%, 380%, 81% higher than OMEN, FLA, PassGAN, VAEPass, and PassGPT respectively.

V. CONCLUSION

In this paper, we recognize that the current neural network password guessing model faces the common problem of high repetition and disordered probability in generating passwords, causing low efficiency in subsequent password attacks. Out of ordinary, we don't focus on modeling but the generation method. We propose a search-based ordered password generation method called SOPG, which enables the autoregressive neural network password guessing model to generate passwords in approximately descending order of probability. Evaluation results show that the performance of SOPG is superior to random sampling comprehensively and our proposed model SOPGesGPT using SOPG to generate passwords gains the state-of-art performance in the test. During this research, we also recognize many of the generated passwords have appeared in the training set, that is another inherent problem in password guessing using deep learning technology and especially severe in the initial stage of SOPG. Next, we will focus on solving this problem.

ACKNOWLEDGMENT

The authors are grateful to the anonymous reviewers for their invaluable comments that guide the authors to revise and improve this paper.